\documentstyle[editedbook,epsfig,psfig,epsf]{mq}
\def\cm2{cm$^2$ }
\def\se1{s$^{-1}$ }

\begin{opening}
\title{Modelling Oscillations in GRS 1915+105 Jet Emission}
\author{R.S. Collins$^{1,2}$, C.R. Kaiser$^1$ \& S.J. Cox$^2$}
\institute{$^1$ Department of Physics and Astronomy, University of Southampton,\\ Hampshire, SO17 1BJ, United Kingdom\\
$^2$ Department of Electronics and Computer Science, University of Southampton,\\ Hampshire, SO17 1BJ, United Kingdom}
\end{opening}

\runningtitle{Oscillating Jet Emission Model}
\runningauthor{Collins, Kaiser \& Cox}

\begin{document}
\vspace{-0.5cm}
\begin{abstract}
{\small We have studied time variability in the flux from the flat spectrum synchrotron radiation of the Blandford \& K\"{o}nigl (1979) model for relativistic, conical jets. The resulting model has been applied to the flux variation of the flat spectrum of GRS 1915+105 observed by Fender \& Pooley (2000). This comparison has highlighted a fundamental problem of the Blandford \& K\"{o}nigl (1979) model in that it requires unphysically large electron densities to explain the flux levels observed from the flat spectrum of GRS 1915+105.}
\end{abstract}

\section{The Model}

To explain the infrared and millimetre emission oscillations from GRS 1915+105, we have adopted the Blandford \& K\"{o}nigl (1979, hereafter BK79)\cite{BK79} model of a relativistic jet, containing power-law electron distributions. Adiabatic expansion gives the jet a conical shape with an electron density and magnetic field strength that decreases with radius. The jet emits partially self-absorbed synchrotron emission over a radial range covering several orders of magnitude, and hence from a wide range of optical depths. Therefore, the total jet emission consists of the summation self-absorbed synchrotron spectra from regions of the jet of decreasing optical depth (or increasing radius), and hence electrons at each radius, from the base to the top of the jet, emit a spectrum that is progressively shifted towards lower frequencies. This summation results in a `flat' region to the spectrum, which has a slope of zero if energy losses due to adiabatic decompression are ignored (as in the BK79 model). However, if some energy losses do occur then this region has a positive, inverted slope, the magnitude of which depends upon the magnitude of the energy loss.

The extent of the flat spectral region, $\nu_{\rm max}/\nu_{\rm min}$, is solely determined by the emission region size, which in our model is fixed by the jet velocity and time-scale of variability. At the lower frequency end of the flat spectrum there is a smooth transition to an optically thick synchrotron spectrum with $F_{\nu} \propto \nu^{5/2}$, and the high frequency end smoothly terminates into an optically thin, $F_{\nu} \propto \nu^{-5/8}$, synchrotron spectrum.

Full detail of the derivation of our model may be found in Collins, Kaiser \& Cox (2002)\cite{Collins02}. The emission spectrum predicted by this model is given in terms of the optical depth function, $\tau_{\nu}(r)$.
\begin{equation}
\label{eFN}
F_{\nu} = \frac{8.8 \times 10^{-18}}{({D_{\rm j}}/ \mbox{pc})^2} 
  \; b_0 \left(\frac{\nu}{\rm GHz}\right)^{5/2} \nonumber
  \int_{r_{\rm min}}^{r_{\rm max}}\left(\frac{r}{r_0}\right)^{3/2} 
  \left[1 - \mathrm{e}^{-\tau_{\nu}(r)}\right] \,\mathrm{d}r \mbox{ mJy},
\end{equation}
where, 
\begin{equation}
\label{eDepth}
\tau_{\nu}(r) = 1.5 \times 10^{-12} a_0 \left(\frac{r}{r_0}\right)^{-25/8}         \left(\frac{\nu}{\rm GHz}\right)^{-25/8},
\end{equation}
$D_{\rm j}$ is the distance to the jet, $r_{\rm max} / r_{\rm min}$ defines the extent of the emission region, and all parameters have SI units unless stated otherwise.

The model has two parameters. The first, which we have named $a_0$, controls the optical depth as a function of radius, $\tau_{\nu}(r)$, and thus shifts the flat region of the spectrum to higher or lower frequencies. The second model parameter, which we have named $b_0$, only effects the flux normalisation. The $a_0$ parameter also has an effect upon the flux normalisation, and therefore we first fit the $a_0$ parameter to the spectral shape, before fitting $b_0$ to the flux normalisation. The model parameter, $a_0$, is related to the physical parameters of the system by $a_0 = w(r\!=\!r_0)k(r\!=\!r_0)B(r\!=\!r_0)^{17/8}$, where $k$ is the normalisation value of the electron density distribution, and $B$ is the magnetic field strength. The $b_0$ parameter is defined as $b_0 = w(r\!=\!r_0)B(r\!=\!r_0)^{1/2}$. The half-width of the jet $w(r\!=\!r_0)$ is fixed by observational constraints, and therefore from $a_0$ and $b_0$, we may determine $k(r\!=\!r_0)$ and $B(r\!=\!r_0)$.

To implement time dependence in this model as a variation in the flux level of the flat spectrum it would be simplest to make $b_0$ a function of time with $a_0$ fixed. However, physically this would require the electron density to vary by a process that exactly compensates variations in the magnetic field strength. Therefore we choose to make $a_0$ a function of time, and fix $b_0$, which may be simply interpreted, physically, as a variation to the electron density injected into the jet. Time dependence is then included into equation \ref{eFN} by defining $r_{\rm max} = v_{\rm j} t$, where $v_{\rm j}$ is the bulk velocity of the jet material, and by increasing the injected value of $k_0$ with time according to a Gaussian function.

From this model we can predict three observable parameters to which our model parameters may be fit; the time lag between the peak fluxes at each frequency of observation, the flux ratio of these peak fluxes, and the flux normalisation value. The value of the $a_0$ model parameter affects all three of these observables. The flux ratio is determined by the position of the flat spectral region with respect to the two frequencies of observation, which is defined by $a_0$. The time lags are also determined by the position of the flat spectral region, because the optically thin end of the flat spectrum is formed at small radii, whereas the optically thick end is formed at much larger radii. Therefore the closer that both frequencies of observation are to the optically thin end,  the smaller the time lag.

\section{Application to GRS 1915+105}

Fender \& Pooley (2000)\cite{FP00} observed large amplitude oscillations in the emission from GRS 1915+105 at an infrared wavelength (2.2 $\mu$m) and a millimetre wavelength (1.3 mm). The time lag between the peak fluxes from each wavelength for the first oscillation was approximately 25 s, with the dereddened infrared flux reaching 525 mJy and the millimetre flux reaching 340 mJy, giving a peak flux ratio of 1.5. However, the extinction correction applied to the infrared flux is uncertain, such that the peak flux ratio is $1.5 \pm 0.6$. Such a small flux ratio over a frequency range covering three orders of magnitude is suggestive of the `flat' spectrum produced by relativistic jets. 

For the fixed parameters of the model we assume that the distance to GRS 1915+105 is 11 kpc, the opening half-angle of its jet is $4^{\circ}$, the base of the emission region is at $r_{\rm min} = 10^5$ m, and the jet material has a bulk velocity of $v_{\rm j} = 0.6 c$. The value of the $a_0$ parameter was fit to the observed time lag, and, for this value of $a_0$, $b_0$ was fit to the flux normalisation of the peak millimetre flux. From these model parameters the corresponding physical parameters of jet were determined to be $B(r\!=\!r_{\rm min}) = 7.8 \times 10^{-7}$ T and $k(r\!=\!r_{\rm min}) = 5.4 \times 10^{40}$ m$^{-3}$. The model also predicts a flux ratio of 1.1 which agrees with the observations.

We have found that it is impossible to obtain flux levels of the order of $\sim 100$ mJy without unphysically high electron densities, for time lags of less than 300 s, as demonstrated by figure 1. This is not just a problem with this time dependent model, but is a fundamental problem of the BK79 model, as can be seen by performing the same fitting process to the observed flux ratio.

\begin{figure}
\centering
\psfig{file=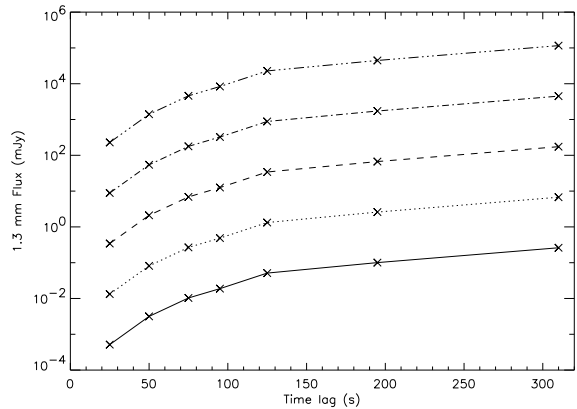,width=8cm}
\caption{Each line in this plot represents the 1.3 mm flux level versus time lag relationship for a different value of the peak injected electron density from the bottom line of $k(r\!=\!r_{\rm min}) = 10^{16}$ (solid line), $10^{22}, 10^{28}, 10^{34}, $ to $10^{40}$ (dashed line) m$^{-3}$ at the top. It is clear that for short time lags (from small values of $a_0$) it is impossible to achieve the observed flux densities without extremely high values for the electron density. Taken from Collins et al. (2002).}
\label{}
\end{figure}

To agree with the observed flux levels and the time lag between peaks with realistic electron densities, the model must either predict a flux $\sim 10^4$ times larger with the fitted value of $k_0$, or predict the observed time lag with a value of $a_0$ that is $\sim 10^9$ times larger. Since relativistic effects are negligible in the GRS 1915+105 jet (see Collins et al. 2002), the later case can be immediately ruled out, as it would require the model's fixed parameters to be adjusted beyond physical limits. It is also difficult to justify the higher flux required by former solution, as Doppler beaming and geometrical effects are not sufficient.

In conclusion it is impossible to explain synchrotron emission of this strength from a BK79 type jet with justifiable electron densities, when restricted to the observed time lags.

\end{document}